
%
%

\magnification=1200
\baselineskip=16pt
\rightline{ILL-(TH)-94-4}
\rightline{Feb. 1994}
\vskip1truecm

\centerline {\bf  LOOKING FOR THE LOGARITHMS IN FOUR-DIMENSIONAL}
\centerline  {\bf NAMBU-JONA-LASINIO MODELS}
\vskip1truecm

\centerline {Seyong KIM}
\centerline {\it High Energy Physics Division, Argonne National Laboratory}
\centerline {\it 9700 S. Cass Avenue, Argonne, Il 60439}
\vskip5truemm

\centerline {Aleksandar KOCI\' C and John KOGUT}
\centerline {\it Loomis Laboratory of Physics, University of Illinois}
\centerline {\it 1110 W. Green St., Urbana, Il 61801-3080}
\vskip3truecm

\centerline {\bf Abstract}
\vskip 0.3truecm
We study the problem of triviality in the four dimensional
Nambu-Jona-Lasinio model with discrete chiral symmetry
using both large-$N$ expansions and lattice simulations.
We find that logarithmic corrections to scaling appear in the equation of
state as predicted by the large-$N$ expansion.
The data from $16^4$ lattice simulations is sufficiently accurate to
distinguish logarithmically trivial scaling from power law scaling.
Simulations on different
lattice sizes reveal an interesting interplay of finite size effects and
triviality. We argue that such effects are qualitatively different
for theories based on fundamental scalar rather than fermion fields.
Several lessons learned here can be applied to simulations and analyses of more
challenging field theories.

\vfill\eject

\noindent {\bf 1. Introduction}
\vskip 0.3truecm

The Nambu-Jona-Lasinio (NJL) model [1] describes
the dynamics of massless fermions that
have a contact interaction.
The theory has two phases. At weak couplings fermions are
massless, while
at strong couplings they acquire dynamical mass due to spontaneous
chiral symmetry breaking.
The transition from a massless to a massive phase
is accompanied by the appearance of light
scalar (and pseudoscalar) bound states. Through the exchange of these
composites an effective interaction develops over physical scales and
the cutoff disappears from the model [1]. The original theory,
although not renormalizable in perturbation theory, becomes
renormalizable and appears to have the same low energy limit as
a linear $\sigma$-model [2]. Since the effective interaction
is of the Yukawa type, it is important that, measured in
physical units, the mesons retain a finite extent in order to ensure
an interacting low energy limit. This is possible
to achieve below four dimension [3,4], but for $d>4$, the mesons are always
pointlike in the continuum limit and the corresponding theory is
free.
In the marginal dimension, $d = 4$, there are calculations and analyses
supporting triviality as well [3,5], but a rigorous demonstration is
lacking.

The reason for triviality in the NJL model is
different from that in the more familiar and more extensively studied
case of $\phi^4$ theory [6]. The essence of the chiral phase transition is
to tune the coupling in such a way to allow for the formation of
bound states. Once bound states appear in the spectrum, everything else
related to chiral symmetry breaking follows naturally [7].
Keeping the size of the bound state finite, on the other hand, is a matter of
balancing the attraction due to the interaction
and the zero-point repulsion due to the kinetic
energy. Above four dimension, this is not possible to achieve
since the short-distance
attraction becomes progressively stronger with increasing
dimensionality and collapsing bound states always lower the energy.

In four dimension, however, the attraction and the repulsion scale identically
and the physics is driven by whatever is left over after their cancellation.
So, the problem of triviality in four dimension amounts to more than
simple power counting and a decisive answer requires a careful study.
The technical difficulties lie in the fact that power counting in
the four dimensional case is modified by logarithmic corrections
and these logarithms
are notoriously difficult to establish both theoretically and
numerically. There are several techniques, such as large-$N$,
$\epsilon$-expansions or perturbation theory, that can provide some hints of
how the logarithms enter the game [3,8,9].
In the  NJL model the physics of triviality
is intrinsically nonperturbative due to the presence of bound states and
chiral symmetry breaking -- the weak coupling phase is symmetric and the
Goldstone physics is not accessible to perturbation theory. Thus,
the only analytic method that can be of some utility in the context
of NJL models is
the $1/N$ expansion which we will use as a guide in analyzing the data.

In the past the NJL model has captured considerable attention
and has stimulated
activities in a wide range of contexts, from purely theoretical [3,10-12]
to practical model building, such as
nuclear physics and the intermediate energy sector of $QCD$
[13,14], technicolor [15]
and the top condensate model [16], to mention a few.
Most of the studies have been performed in the large-$N$ limit,
for the reasons mentioned above, and systematic studies of the $1/N$
corrections have been carried out. The $1/N$ expansion indicates that
below four dimension the theory is renormalizable [3,8], whereas for $d\geq4$
it is trivial. There is little doubt that this changes at low-$N$, but
no proof outside of the $1/N$ expansion exists.
In that sense we have undertaken this project in order to gain some insight
and develop some criteria to find logarithms in a
theory that is known to have them.

Recently several groups
have carried out numerical simulations of various versions of the
NJL model, at [16] and below four dimension [18,19],
but the problem of triviality
in four dimension has not been tackled intensively before.
In this paper we want to address precisely this issue with an emphasis on
how the logarithms should be tracked down. We should point out
that we do not expect any surprises as far as the large-$N$ predictions are
concerned and, in that sense, we anticipate the outcome that the theory is
trivial. Our main motive is to see, knowing the answer,
with what degree of confidence we can establish
it numerically. Our main efforts will be concentrated
on the equation of state (EOS) and the logarithms that accompany critical
scaling. We have been encouraged to undertake this project by the recent
observation that the logarithms in models with composite mesons,
(therefore the NJL model), show up in a {\it qualitatively} different way
than in theories with elementary scalars, like $\phi^4$ [20].
Interestingly, the logarithmic behavior from our simulations on
large lattices ($16^4$) agrees with the composite meson case,
as expected. However the
logarithmic behavior from the simulations on the smaller lattices ($8^4$)
is consistent with the elementary scalar case.
This apparent disagreement between two lattice simulation results must
be due to the systematics of finite size effects and understanding
the reason behind it will increase our confidence in deciding
whether a particular lattice result is tainted by finite size effects.
Our efforts include some technical aspects of this problem as well.
In particular, we will show
how to distinguish power law scaling from mean field scaling modified
by logarithmic corrections. This problem is ``reversed"
from searches for a nontrivial fixed point in other models.
As will be discussed in section 4, when we discuss the lattice data,
the logarithms in the NJL model reveal themselves
in a clear and unambiguous way. Especially interesting, in this context, is
the manner in which the power law hypothesis fails.

There are important lessons to be learned
from this exercise alone. They concern numerical simulations of theories with
nontrivial fixed points for which no analytic tools are available.
On a technical level, establishing the logarithms (therefore triviality),
is a more demanding job than diagnosing their absence in theories
that are not trivial.

Setting our goal as ``logarithm hunting", we have
chosen to simplify our task as much as possible. For that reason, we decided
to work with $N=12$ flavors in order to remain close to the large-$N$ limit.
This turned out to be a profitable decision for another reason.
Tunneling is suppressed at large-$N$, so we could work relatively near
the model's critical point without suffering from
this particular finite size effect.
Also, the choice of discrete chiral symmetry, instead of continuous
chiral symmetry
enables us to work in the chiral limit and avoid the extrapolations
to zero bare mass which complicate other studies in the field.

The layout of this paper is as follows. In Sec.2 and 3 we solve the model in
leading large-$N$ order and identify the logarithms of triviality
in the equation of state, susceptibility, and specific heat. These
results are contrasted to the analogous quantities in $\phi^4$. In
Sec.4 we present the simulation data and fits on $16^4$ and
$8^4$ lattices. We shall see that the $16^4$ lattice simulations
expose the logarithms of the large-$N$ expansion with considerable
precision. The $8^4$ data is strongly distorted by finite size effects
which in some cases obscures the logarithms in curious ways. In Sec.5
we discuss our results and the lessons we have learned which can be
applied to studies of more challenging field theories.

\vskip5truemm
\noindent{\bf 2. Large-$N$ limit }
\vskip3truemm

In this and latter sections we review the large-$N$ limit of the NJL model.
The emphasis will be on the scaling violations and how they enter
the EOS and expressions for other thermodynamic quantities.
In order to facilitate the large-$N$ expansion,
the four-fermi lagrangian can be written in the form of a Yukawa theory

$$
L=\bar\psi (-i{\not\! \partial} +m+g\sigma )\psi -{1\over 2}\sigma^2
\eqno(2.1)
$$
The original, four-fermi, lagrangian is recovered after integrating
over the $\sigma$ field.
Since $L$ is quadratic in fermionic variables, we can integrate them out
and obtain the effective action

$$S_{eff}={1\over 2}\int_x\sigma^2 - Ntr\ln (-i{\not\! \partial}+m +g\sigma)
\eqno(2.2)
$$
In the large-$N$ limit the dominant configurations are obtained by
the saddle point method. The resulting equation of state (EOS) is
the minimum condition $\sigma -Ngtr\int_q 1/({\not\! q}+m+g\sigma )=0$.
After identifying the fermion mass as $\Sigma=m+g\sigma$,
the minimum condition reads

$$
\Sigma = m -g^2<\bar\psi\psi>
\eqno(2.3)
$$
The curvature of the effective action is the inverse susceptibility

$$
{ {\delta^2 S_{eff}}\over{\delta\sigma^2} }=\chi^{-1}= 1
-4g^2\int_q{ 1\over{q^2+\Sigma^2} }+ 8g^2\int_q{ {\Sigma^2}
\over{(q^2+\Sigma^2)^2} }
\eqno(2.4)
$$
The appearance of soft modes at the critical point
is manifested in the divergence of the susceptibility and the
vanishing of the fermion mass.
Setting $\Sigma=0$ in eq.(2.4) gives
$1-4g_c^2\int_q 1/q^2=0$.
With this expression for $g_c$, the EOS can be recast in the form

$$
{m\over\Sigma} +t=
4 g^2\int_q{ {\Sigma^2}\over{q^2(q^2+\Sigma^2)} }
\eqno(2.5)
$$
where $t=g^2/g^2_c -1$. This is a convenient form for extracting
the critical indices since one need only count the infrared divergences
of the right hand side. The indices associated with the EOS are defined as
[21]:
$<\bar\psi\psi>\sim t^\beta$ and  $\chi \sim t^{-\gamma}$ at $m=0$, and
$<\bar\psi\psi>\sim m^{1/\delta}$ at $t=0$.
Since there are only two fields the system responds to, only two
out of three indices are
independent. They are related by the scaling relation $\beta(\delta-1)=\gamma$.
Above four dimension, the integral in (2.5) is infrared
finite, and the exponents do not change with dimensionality -- they have
the mean field values: $\beta=1/2, \delta=3, \gamma=1$. Below four dimension
the right hand side scales as
$\Sigma^{d-2}$ resulting in nongaussian critical behavior
with indices $\beta=1/(d-2), \delta=d-1, \gamma=1$.
In four dimension, the integral is
logarithmically divergent giving rise to scaling violations.
The EOS in this case reads

$$
{m\over\Sigma} +t=
ag^2 \Sigma^2 \ln {b\over{\Sigma^2}}
\eqno(2.6)
$$
where $a$ and $b$ are constants, and the cutoff is set equal to one.
Using a momentum cutoff gives
$a=1/4\pi^2$ and
$b=1$. In the chiral limit, the fermion mass scales as the order parameter,
$\Sigma \sim <\bar\psi\psi>$.
At the critical point
$\Sigma =(m+C m^{1/\delta})$, but since $\delta >1$, the linear
term is subleading and the scaling of $\Sigma$ and $<\bar\psi\psi>$
coincide again. The two limits of the gap equation give

$$
t=ag^2 \Sigma^2 \ln {b\over{\Sigma^2}}\,\,\,\,\,\,\,\,
(m=0)
\eqno(2.7a)
$$
$$
m=ag^2 \Sigma^3 \ln {b\over{\Sigma^2}}\,\,\,\,\,\,\,\,
(t=0)
\eqno(2.7b)
$$
Up to logarithms the exponents have their gaussian values $\beta =1/2$ and
$\delta =3$.

There are two universal quantities to be extracted from the
susceptibility. In the critical region, in the chiral limit,
the susceptibility scales as $\chi\approx C |t|^{-\gamma}$. The
amplitude $C$ is different in the two phases and the ratio $C_+/C_-$
is a universal quantity. Mean field theory predicts $C_-/C_+ =2$ [21].
In the symmetric phase it is easy to compute both $C_-$ and $\gamma$.
Setting $\Sigma=0$ in eq.(2.4) and using the definition of the
critical coupling, we get

$$\chi^{-1}=1-g^2/g^2_c=(-t)
\eqno(2.8a)
$$
which implies $\gamma=1$, $C_- =1$. In the broken phase
the first two terms in eq.(2.4) cancel because of the gap equation.
Combining the third term in eq.(2.4) with the gap equation (2.5),
gives $\chi^{-1} -2t\sim \Sigma^2$and the expression for the susceptibility
becomes

$$
\chi^{-1}=  \biggl(2-{ 2\over{\ln(b/\Sigma^2)} }\biggr) t
\eqno(2.8b)
$$
Clearly, the exponent $\gamma=1$ is the same as in the symmetric phase.
There are no logarithmic corrections to scaling for the susceptibility.
The amplitude, however, receives corrections due to scaling violations so that

$$
{ {C_-}\over{C_+} } =2-{2\over{ \ln(b/\Sigma^2)} }
\eqno(2.9)
$$


The wavefunction renormalization constant is given by
the $k^2$ term of the $\sigma$ propagator. Due to radiative corrections
such a term is generated in the $1/N$ expansion.
The leading log contribution is given by

$$
Z^{-1}= 2g^2\int_q{ 1\over{(q^2+\Sigma^2)^2} }\sim \ln {1\over{\Sigma^2}}
\eqno(2.10)
$$
The anomalous dimension $\eta$, defined by $Z\sim \Sigma^\eta$, vanishes
up to logarithms.

There are basically two candidates for the correlation length in this model:
the fermion and $\sigma$-masses.
It is simple to verify that they both scale the same way.
The mass of the $\sigma$ particle, for example,
is obtained through the second moment $\chi$. Using eqs.(2.8)
and (2.10) leads to $M^2=Z\chi^{-1}\sim \Sigma^2$
The scaling of the fermion mass, $\Sigma$,
is obtained easily from the
gap equation, (2.7a). It follows that $\nu=\beta$, and so
$\nu = 1/2$ with the same scaling violations
as for the order parameter.


The specific heat is obtained from the effective action, eq.(2.2).
Like the  susceptibility, it too is
characterized by two quantities, the exponent
$\alpha$ and amplitude $A$.
In the critical region its scaling is given by
$C=A|t|^{-\alpha}$. The
amplitudes are different in the two phases and their ratio $A_-/A_+$
is universal [21]. Mean field theory predicts $\alpha=0$, $A_-=0$
and the specific heat undergoes a jump. Unlike other exponents,
$\alpha$ can be either positive or negative. In four dimension
logarithmic corrections affect the scaling of $C$. However, to
leading order in $1/N$, $A_-$ remains zero. It is easy to see this since
in the
symmetric phase, $<\sigma>=0$, and the action is independent of $g$. Thus,
$A_-=0$. In the broken phase, straightforward algebra leads to

$$
C\sim {1\over{\ln(1/t)}}
\eqno(2.11)
$$
Higher order corrections might change the power of the logarithm,
but it remains in the denominator.
For scalar theories there are several results on the amplitude ratio
and critical index $\alpha$ ($\epsilon$-expansion and large-$N$) [22].
In the symmetric phase $A_-=O(1/N)$, so it
does not appear in leading order. The amplitude ratio, therefore,
scales as $1/N$. In four dimension, $\alpha=0$ as well, but with logarithmic
corrections of the form $C\sim 1/(\ln t)^{(N-4)/(N+8)} $. Thus, depending
on $N$, the power of the logarithm can
change it's sign and the specific heat can change
its behavior from vanishing to divergent. The $\epsilon$-expansion studies
indicate that this does not happen in Yukawa models [19].
\vskip5truemm
\noindent{\bf 3. Triviality of the four-dimensional theory
in the large-$N$ limit }
\vskip3truemm

To establish the connection between scaling violations and triviality,
we introduce the renormalized coupling. It
is a dimensionless low-energy quantity
that contains information
about the non-gaussian character of the theory. It is conventionally
defined as [23]

$$
g_R=- { {\chi^{(nl)}}\over{\chi^2 \xi^d} }
\eqno(3.1)
$$
where the nonlinear susceptibility $\chi^{(nl)}$ is the zero-momentum
projection
of the connected four-point function

$$
\chi^{(nl)}={ {\partial^3 M}\over{\partial h^3} }=
\int_{123}<\sigma(0)\sigma(1)\sigma(2)\sigma(3)>_c
\eqno(3.2)
$$
The normalization factors, $\chi = \int_x<\sigma(0)\sigma(x)>_c$ and $\xi^d$,
in eq.(3.1) take
care of the four fields and the three integrations.
In a gaussian theory all higher-point functions
factorize, so $g_R$ vanishes.
Using the hyperscaling hypothesis, this can be converted into,

$$
g_R \sim \xi^{(2\Delta-\gamma-d\nu)/\nu}
\eqno(3.3)
$$
where $\Delta=\beta+\gamma$ is the gap exponent.
Being dimensionless, $g_R$ should be independent of $\xi$ if $\xi$ is the only
scale. Thus, the validity of hyperscaling
requires that the exponent must vanish. It
implies the relation, $2\Delta-\gamma -d\nu=0$,
between the critical indices.
In general, it is known that the following inequality [24] holds

$$
2\Delta\leq \gamma+d\nu
\eqno(3.4)
$$
The exponent  in the expression for $g_R$ is always
non-positive, so that violations of hyperscaling imply that the resulting
theory  is
non-interacting.
Above four dimension, the exponents are gaussian ($\gamma=1,\Delta=3/2,
\nu=1/2$). In this case, it is easy to verify the above inequality:
$3\leq 1+d/2$, which amounts
to $d\geq 4$.
In four dimension there are logarithmic corrections to scaling that
are believed to drive $g_R$ to zero.
Scaling violations in any thermodynamic quantity propagate into the
renormalized
coupling and, according to eq.(3.4), these violations lead
to triviality. Because this is a marginal case, it requires special
treatment.

In order to make our point simple, we
illustrate how logarithms appear in the effective actions of
two solvable models:
$\phi^4$ and $(\bar\psi\psi)^2$ theories both in the
large-$N$ limit.
The effective actions for the two models are [5,25]

$$
V(M)=-{1\over 2}t{ {M^2}\over{\ln(1/M)} }
+{\lambda\over 4} {{M^4}\over{\ln(1/M)}}
\eqno(3.5a)
$$
$$
V(\sigma)=-{1\over 2}t\sigma^2+
\sigma^4\ln(1/\sigma)
\eqno(3.5b)
$$
This is the leading log contribution only.
In the first example, it is clear how log-corrections lead to
triviality. The logarithm can simply be thought of as coming from the
running coupling --
quantum corrections lead to the replacement $\lambda\to \lambda_R$.
The vanishing of the renormalized coupling is then manifest from eq.(3.5a).
In the case of fermions, eq.(3.5b),
the details are completely different -- the analogous
reasoning would lead to an erroneous
conclusion that the renormalized coupling increases in the infrared.
This form is obtained from the effective action,
$S_{eff}(\sigma)$, in eq.(2.3). It is easy to see how the logarithm in (3.5b)
appears in the numerator.
For example, after taking the second derivative
of $\chi^{-1}$,
eq.(2.4), the last term generates a logarithmically
divergent contribution which is the log-term in eq.(3.5b).
The vanishing of the renormalized coupling here follows
from the wave function renormalization constant
$Z\sim 1/\ln(1/\sigma)$ [5].

In both cases the renormalized coupling is obtained through the
nonlinear susceptibility. The correlation length is related to the
susceptibility by $\xi^2=\chi/Z$. For magnets (i.e. scalar theories like
$\phi^4$) the following relations hold,

$$
\chi^{(nl)}\sim \chi^4 {\lambda\over{\ln(1/M)}},\,\,\,\,\,\,\,\,\,\,\,\,
Z=1
\eqno(3.6a)
$$
$$
g_R\sim Z^2 {\lambda\over{\ln(1/M)}}\sim {1\over{\ln(1/M)}}
\eqno(3.6b)
$$
so $g_R$ vanishes at a logarithmic rate.

For fermions, on the other hand, we have

$$
\chi^{(nl)}\sim \chi^4 \ln(1/\sigma),\,\,\,\,\,\,\,\,\,\,\,\,
Z\sim{1\over{\ln(1/\sigma)}}
\eqno(3.7a)
$$
$$
g_R\sim Z^2 \ln(1/\sigma)\sim {1\over{\ln(1/\sigma)}}
\eqno(3.7b)
$$
and again $g_R$  vanishes at a logarithmic rate.
In this context the following point should be made.
The nonlinear susceptibility in the NJL model is a
connected four-point function for the composite $\bar\psi\psi$ field. The free
fermionic theory is not gaussian in $\bar\psi\psi$, so even in free field
theory
$g_R$, defined in this way,
does not vanish. The fact that $g_R\to 0$ near the critical point
indicates not only that the resulting theory is gaussian, but that it is
also a purely bosonic theory. So, for the NJL model, the bare theory and its
continuum limit bear no resemblance.
This is quite different from what
happens in the $\phi^4$ model where the resulting continuum limit has the
same physical content as the bare theory but with vanishing coupling constant.

The important point in this comparison of the two models
is the fact that the logarithms in the
EOS appear in different places. This property is generic for the two models
and persists beyond the $1/N$ expansion. In order to emphasize this point, we
consider the critical EOS for both models.
They are obtained from
the effective potential by simple differentiation.
To make the connection with $\delta$, we take $t=0$.
The critical EOS for the magnets is [25]

$$
h\sim {  {M^3}\over{\log(1/M)}  }
\eqno(3.8)
$$
This defines the exponent $\delta$. Because of the scaling violations,
eq.(3.8) vanishes {\it faster} than a pure power. So the ``effective" $\delta$
is bigger then its mean field value. It is easy to understand the
correction to scaling in eq.(3.8). $1/\delta$ is simply the logarthmic
derivative of the order parameter $1/\delta = \partial\ln M/\partial \ln h$.
Since the magnetization saturates at strong fields, its slope vanishes in the
large-$h$ limit. Thus, $1/\delta$ decreases away from the $h=0$ limit.
As was shown in ref.[20],
such corrections to scaling can never occur in
the case of the chiral transition.

For the NJL model the critical EOS reads,

$$
m\sim   \sigma^3\log(1/\sigma)
\eqno(3.9)
$$
Unlike scalar theories, the log's appear in the {\it numerator} --
the right hand side in eq.(3.9)
vanishes {\it slower} than the pure power and the ``effective" $\delta$
is {\it smaller} than the (pure) mean field value. This is also easy to
understand since at large mass $<\bar\psi\psi>\sim m$. Thus, away from
the chiral limit $1/\delta$ is driven to 1. As pointed out in ref.[20]
this difference in the position of the logarithm, eqs.(3.8-9)
or, equivalently,
the sign of the scaling violations is generic for the two models and is
intimately tied to the fact that in NJL scalars are composite and in $\phi^4$
they are elementary.

The distinction between the two models proved to be useful in analyzing
the lattice data. In particular, if the NJL model is simulated on a {\it small}
lattice, then, for a fixed
mass, the right hand side
of eq.(3.9) will be smaller than it is in the thermodynamic
limit because finite volume effects tend to restore the symmetry,
and reduce the order parameter.
If the infinite volume $\sigma$ is to be fitted to the data in the form
$m\sim \sigma^{\delta'}$, then
the appropriate exponent $\delta'$, which might fit the data,
must be {\it bigger} than its actual value. This larger value of $\delta$
could also be mimicked by mean field scaling supplemented with the
log in the denominator.
Such data should be not be trusted since their behavior is
qualitatively different from the behavior of the system in the
thermodynamic limit. As the
lattice size is increased, the effective $\delta$ would eventually fall
below its mean field value suggesting the proximity of the thermodynamic
limit. Such data, although not necessarily completely free
of finite size distortions, do share the same qualtiative features
with the physics of the infinite volume.
So, for fermions the bound $\delta <3$ separates the data into two groups
and tells us which one is more reliable. This is possible
because finite size distortions and the approach to the true scaling
law have the {\it opposite} tendencies.
For scalar theories, on the other hand,
the two have the {\it same} tendency.
Nontrivial scalar theories have $\delta>3$,
to begin with. For example, from eq.(3.8) it is apparent that
in case of logarithmic corrections to mean field scaling,
the finite size distortions which tend to reduce
the magnetization can be masked
by adjusting either the exponent $\delta$ or
the strength of the logarithm
so that the right hand side of the critical EOS
assumes an appropriate magnitude compatible with the given $h$.
In fact, strong finite size effects would force the fit with
$\delta>3$ even if the scaling is mean field, so work
on a larger lattice is necessary to either confirm or disprove this result.
Thus, in scalar theories
a careful finite size scaling analysis seems to be
necessary to disentangle the log's from the finite size effects.
In that sense, establishing triviality in $\phi^4$ is more delicate than
it is in fermionic theories [6].
\break

\vskip0.5truecm
\noindent{\bf 4. Simulation results}
\vskip0.2truecm

We carried out extensive simulations of the model described by Eq.(1)
on $8^4$ and $16^4$ lattices. The small lattice simulations were done
to monitor finite size effects. Of course, many more
calculations on a variety of lattice sizes could be done to make
our considerations on finite size effects more quantitative,
but $8^4$ and $16^4$ simulations will prove adequate for our purposes.
If Four Fermi models continue to play an ever increasing role in high energy
theory, then more thorough studies would be needed.

Although there are some cases in which the finite size effects
were large and curious as we shall see below, most of the $16^4$
simulations were relatively free from finite size problems for rather wide
ranges of couplings and yielded precise continuum scaling laws. In fact,
the logarithms of triviality will be seen clearly in our study of the
Equation of State. There are several reasons for this success. First, the
model can be simulated for zero quark mass. So we are free of the difficult
issues of extrapolations to the chiral limit. Second, the discrete character
of the symmetry breaking made the Hybrid Monte Carlo algorithm particularly
effective and allowed us to run the simulations with a relatively large
time step (dt=0.1) while still having a high acceptance rate (from 60 to 90
percent was typical). Third, we could dial the number of flavors large
to compare to the $1/N$ expansion (we chose $N_f=12$, as in our past work
in three dimension [18]) which suppresses fluctuations enormously.

As discussed in the theory sections above, scalar systems and fermion
systems are clearly distinguished by the positions of the
scale breaking logarithms in their critical behaviors. One of the
more interesting features we discovered here is that those distinct
behaviors can be blurred by finite size effects. Comparison between
$16^4$ result and $8^4$ clearly illustrates such effects.

The lattice action, the Hybrid Monte Carlo algorithm and other technical
details of the simulations follow ref[18] in detail. The interested
computational physicist should consult that reference for details. In
our discussion and Tables, we shall add any technical information which
was special to working in four dimension.

For clarity, we shall organize our lattice simulation results into
various subtopics, following the theoretical discussion above.

\vskip0.5truecm
{\it Equation of State}
\vskip0.3truecm

In Table I we have collected the $16^4$ data for the average action, its
specific heat, the vacuum expectation value of the $\sigma$ field, and
its susceptibility $\chi$. We also tabulate there the statistics of each
measurement and the statistical errors on each quantity. The errors were
calculated with considerable care using the usual binning techniques,
because we are interested in high quality fits to distinguish different
functional forms in the model's critical behavior. Note that in the
immediate vicinity of the critical point, we accumulated particularly
high statistics ($1/3$ of a million sweeps) to combat critical slowing
down.

We first considered the order parameter's dependence on coupling in the
broken symmetry phase. The logarithmic fit,

$$
\beta-\beta_c=a\sigma^2(\ln(1/\sigma)+b)^p
\eqno(4.1)
$$
was tried for the twelve data points in the coupling range $\beta=.54-.595$.
A very good fit was found (confidence level=70 percent) and the parameters
were determined: $\beta_c=.6044(1)$, $a=2.25(2)$, $b=1.16(5)$, and
$p=1.041(12)$. The data and fit are shown in Fig. 1a.

Perhaps the most interesting feature of this result is the fact that the
power of the logarithm in Eq.(4.1) is determined with such precision and
confidence. The fitting routine (standard least square method)
searched for the best value
of the power and found 1.041(12). Note that the fit does {\it not} just
accommodate the immediate vicinity of the critical point : the parameter `b'
in Eq.(4.1) which sets the scale of the logarithm allows us to fit both
the strong coupling region (where $\sigma$ is large and mean field scaling
should apply) and
the critical region (where fluctuations causes the logarithmic correction
to the mean field behavior of the model).

It is interesting to plot the data in a slightly different
way to show the importance of the logarithm more visually. Consider
Fig. 2a where we plot the data as $(\beta-\beta_c)/\sigma^2$ vs. $\ln(1/
\sigma)$ for the choice $\beta_c=.6044$. The numerical significance
of the logarithm and the fact that its power is very close to 1.00
are apparent.

We also attempted equation of state fits which are free of logarithms. For
example, if the theory had pure power law singularities, then the form,

$$
\beta_c-\beta = a\sigma^{1/\beta_{mag}}+b\sigma^2
\eqno(4.2)
$$
should apply in the broken phase. When a least square fit of this form
is attempted, the iterative procedure ( the search for the best confidence
level ) does not converge, although high confidence levels are found
for fixed fitting parameters. The fitting program finds that it can
always increase the confidence level by increasing `a' positively, while
increasing `b' negatively and driving $1/\beta_{mag}$ to 2 from above.
Such a tendency is reminiscent of the $\epsilon$-expansion where the
limit of $d=4$ is marked by the fact that power law scaling is converted into
logarithms through the formula
$\ln x = \lim_{\epsilon \to 0} (x^\epsilon  -1)/\epsilon$.
The behavior of the fit is compatiable with the weaker logarithmic dependence
on $\sigma$ found above. Our data is sufficiently accurate and it
extends over a wide enough range of $\beta$ to favor the logarithmic
equation of state over the pure power law formula.

In Table II we collect our $8^4$ measurements which parallel the
$16^4$ measurements presented in Table I. First consider the
equation of state in the broken phase. Fits of the form Eq (4.1) and
(4.2) which should have worked in both the strong coupling and critical
regions were unsuccessful ( the confidence levels were infinitesimal ).
However, if we took a subset of the data in the critical region alone
reasonable fits could, of course, be made. For example, a pure power law
fit such as,

$$
\sigma = a(\beta_c-\beta)^p - b
\eqno(4.3)
$$
could be made over the relatively narrow
region from $\beta = .57$ to $.595$ with a
fair confidence level (20 percent) and results: $\beta_c = .607$,
$a = 1.56(9)$, $p = .54(3)$, and $b=.002(13)$. The fit is shown in
Fig. 1b. Note that the fit is consistent with mean field theory,
unadorned with logarithms.

To gain more insight into the problems with $8^4$ simulation data,
let us plot it as we did the $16^4$ data in Fig. 2a. The result is
shown in Fig. 2b. Comparing with Fig. 2a, it appears that the $8^4$
data suffers from finite size effects as $\beta$ approaches
the critical point
and there are only `hints' of the true logarithmic
corrections in Fig. 2b. This plot shows clearly that it would not
be sensible to `blindly' fit $8^4$ data to any continuum model
scaling law, even though the $16^4$ data shows continuum behavior over
the same range of couplings.

\vskip0.3truecm

{\it Susceptibility}
\vskip0.3truecm

The measurements of the order parameter $\sigma$ allow us to compute
its variance and construct the model's longitudinal susceptibility
$\chi$. The data is given in Table I. First consider the broken
symmetry side of the transition and a determination of the critical
index $\gamma$ assuming a pure power law singularity. For $\beta$
ranging from .54 through .595, we fit the susceptibility data with

$$
\chi^{-1}=a(\beta_c-\beta)^{\gamma} + c
\eqno(4.4)
$$
We are particularly interested in fits which produce an estimate of
the parameter `c' consistent with zero, indicating that $\beta_c$
is a critical point. The least square routine gives: $a=9.14(1.12)$,
$\gamma = 0.967(45)$, $c = 0.002(10)$ for the choice $\beta_c = .6028$.
If trial values of $\beta_c$ are chosen larger than .6045 or smaller
than .6022, then the parameter `c' is pushed significantly away
from zero. However, for {\it all} such choices, $\gamma$ is found to be
.967(45). So this result is very robust given a power law hypothesis.
The data and the fit are shown in Fig. 3a. the confidence level of the fit
is 63 percent (chi-squared = 7.13 with nine degrees of freedom).

As discussed in the theoretical sections above, there are weak logarithmic
corrections to power law scaling of the susceptibility in the broken
phase. So, it is interesting to try fits of the form,

$$
\chi^{-1}=a\biggl(2-{1\over{\ln(1/\sigma)+1.16}}\biggr)
(\beta_c-\beta)^{\gamma}+c
\eqno(4.5)
$$
where the scale in the logarithm (i.e. 1.16) has been taken from
the fit Eq.(4.1).
Again, very good fits are found. Requiring that `c' be compatiable
with zero, produces the prediction $\beta_c = .6040(8)$, $\gamma = 1.054(47)$,
and $a = 7.41(95)$. Although $\beta_c$ is not predicted as accurately as
in the Equation of State fits, the critical index $\gamma$ is again
found for {\it all} choices of $\beta_c$ to be 1.054(47). The confidence
levels of these fits are typically 55 percent. Clearly the logarithm in
Eq.(4.5) is not very significant numerically due to the presence
of the substantial additive term 2. If we had a wider range of $\sigma$
values, the logarithm would play a greater role. The logarithms of
triviality are much more easily exposed in other observables of this
model.

Now consider the susceptibility on the symmetric side of the transition. No
logarithms are expected in the scaling law here, so we try a fit of the
form Eq.(4.4). We find excellent fits (confidence levels of 99 percent)
with : $a = 5.62(1.65)$, $\beta_c = .6024(7)$, and $\gamma = 1.0(1)$.
Again, $\gamma$ is compatiable with unity, as expected theoretically at
large N. The fit is shown in Fig. 3a.

For the $8^4$ lattice, we accummulated susceptibility data in both the strong
and weak coupling phases, and we attempted fits of the
form Eq.(4.5) that were successful on the $16^4$ lattice.
To avoid vacuum tunnelling we did not study couplings
too close to the critical point. The ranges of $\beta$ = .57-
.595 and .62-.655 satisfied this criterion. The data and power law
fits are shown in Fig. 3b. In the broken phase the fit gave
a critical index $\gamma$ = .97(12), compatiable with
theoretical expectations. However, the confidence level of
the fit is barely acceptible (9.6 percent corresponding to a
chi-squared of 6.4 with 3 degrees of freedom), and the other
parameters in the fit were a = 11.3(3.7) and b = $ -.029(29)$
for a critical coupling of $\beta_c$ = .6066.
Note that the crude strong coupling data shown in the figure
at $\beta$ = .54, .55, and .56 were {\it not} used in the fit. The
weak coupling data and its fit are also shown in the figure. The
confidence level was very good (93.9 percent) and the
fit predicted the critical
index $\gamma$ = 1.21(10) with a = 9.8(2.8) and b = .025(8) for
$\beta_c$ = .6066. So the fitted $\gamma$ is two standard
deviations from the theoretical prediction. Since the $16^4$
fits work beautifully with $\gamma$ = 1.05(5), this discrepancy
is most probably a finite size effect.

In summary, the $8^4$ susceptibility study is
not quantitatively reliable,
in contrast to the $16^4$ results which are very
close to the continuum model predictions.

\vskip0.5truecm
{\it Equation of State at Criticality}
\vskip0.3truecm

Another feature of the theory which displays the logarithm of triviality
in an accessible expression is the equation of state at criticality. In
Sec.3 above we contrasted the behavior of scalar vs. fermion models in
Eq.(3.8) and Eq. (3.9). To distinguish the two behaviors we must simulate
the model at $\beta_c$ in the presence of explicit symmetry breaking. We
anticipate that this work will be subject to large finite size effects
and will not be as quantitative as our other measurements which avoided
all the pitfalls of the critical point. We shall see that our results
will favor Eq.(3.9) over Eq.(3.8), and a comparison with $8^4$ measurements
will expose some particularly interesting finite size effects.

We ran simulations at nonzero fermion mass m at $\beta = .6034$. The results
are compiled in Table III. It would be interesting to repeat this work
at other estimates of $\beta_c$, but high statistics are needed since we
are working very close to criticality and statistical fluctuations
are large. We begin by plotting the data in Fig. 4a ($m/\sigma^3$ vs.
$\ln(1/\sigma)$)
in order to discriminate between scalar and fermion behavior. The figure
shows a `window' in $\sigma$ where the logarithm of triviality is in
the numerator of Eq.(3.9). However, it  appears that only a small range
of $\sigma$ is relevant : for large m (greater than .015) the scaling
form of Eq.(3.9) probably does not apply, and for
small m (less than .002) finite
size effects are probably distorting the $16^4$ data. We will gain a better
perspective when the $8^4$ data is plotted. The solid line showing logarithmic
behavior is just meant to guide the eye --- systematic fitting procedures
are not appropriate here.

We also measured the susceptibility at criticality. It follows from
eq.(2.4) that,

$$
\chi\sim {1\over{\sigma^2 \ln(1/\sigma)}}
\eqno(4.6)
$$
so a plot of $\chi^{-1}/\sigma^2$ vs. $\ln(1/\sigma)$ should be linear with
a positive slope. Fig.5 shows that the data is compatiable with this
expectation, but the data suffers from large error bars which make precise
predictions meaningless here.

The $8^4$ data in this case will be subject to much larger and
interesting finite size effects. The data is shown in Table IV
for an estimate of $\beta_c$ = .6066 which follows from the
equation of state and susceptibilities. Masses ranging from
.001 through .020 were simulated, and the results are plotted
in Fig. 4b in the form
suggested by Eq.(3.9). We find that the plot of $m/\sigma^3$
vs. $\ln(1./\sigma)$ has a negative slope! This should be compared
with Fig.4a, where an identical plot was made on the larger lattice
$(16^4)$. The two plots differ qualitatively.
In fact, if we plot
$\sigma^3/m$ vs. $\ln(1./\sigma)$ as suggested by Eq.(3.8), the
scalar field prediction, then a reasonable plot shown in Fig.6
follows. This bizarre situation shows that finite size effects
distort the true behavior of Eq.(3.9), as found on the $16^4$
lattice, into something compatiable with the scalar field
expectation. This should be a severe warning for other
simulation studies : finite size effects can even obscure the
fundamental fermion and/or scalar character of the underlying theory.

\vskip0.5truecm

{\it The Average Action and the Specific Heat}
\vskip0.3truecm

Since accurate measurements of the average action and its associated
specific heat are relatively easy, let us discuss them briefly. The data is
taken from Table I, and the average action is plotted vs. $\beta$ in
Fig.7a. We see a clear change in slope near $\beta \sim .60$. This effect
is clearer in the specific heat which is plotted in Fig.8a. As discussed in
the theory sections above, mean field theory predicts a jump in the
specific heat at the critical point. Possible curvature and logarithms
in such plots are not visible given the error bars in the figures and
the fact that we could not simulate the model closer to $\beta_c$ due
to finite size effects such as vacuum tunnelling. Nonetheless, the
(chiral) phase transition is apparent in the plots.

The $8^4$ data and plots for the average action and its
associated specific heat are given in the table II and Fig. 7b and
8b. These results are qualitatively similar to the $16^4$
results discussed above. They were not useful in subtle studies
of logarithms.
The general agreement between the data from the two lattice sizes
is not unexpected since these quantities are dominated by
ultraviolet fluctuations.

\vskip0.5truecm
\noindent{\bf 6. Discussion and conclusions}
\vskip3truemm

The main conclusions of our study are the following.
Logarithmic corrections to mean field scaling in the EOS
are properly predicted by the large-$N$ expansion.
Our data demonstrate this with a high degree of
confidence. In fact, when we tried to impose the only remaining possibility,
power law scaling in the form of eq.(4.2), not only did
the fit fail, but it did so in an illuminating fashion -
the parameters of the
fit flowed toward a
polynomial generator for the logarithm itself. A pure mean field form
worked only if a restricted range of data were used.

We have seen that finite size effects appear in a peculiar fashion
and we explained the criteria that can detect the finite volume distortions
when they are present.
These criteria are of broader
significance since they originate from studies of the simplest of the
models in the class -- NJL with discrete chiral symmetry.
Other models with dynamical symmetry breaking, like
NJL with continuous chiral symmetry or massless fermionic $QED$, are
subject to more severe
finite volume effects because of their Goldstone bosons.
In that context, the most striking feature of
our study can be summarized in the comparison of the figures 4a and 4b.
They depict identical plots on two different lattice sizes and, because
of the way the plots are designed, the difference is seen as
qualitative. Theoretical
input, layed out in ref[20] and summarized in Sec.3,
would have enabled us to discard the $8^4$ data {\it even if we
had had no access to the $16^4$ lattices}.
While in
some instances, the $8^4$ data were not severely distorted (e.g. Fig.3b),
in other instances, they were completely useless even as a guide for
qualitative behavior. In fact, the logarithms and finite size effects
in these cases interfere with each other and the data is misleading.
Thus, on a small lattice, some fits would favor the logarithms in places
where they simply can not appear.

As it turned out, establishing the log's in the NJL model was easier than
the corresponding task in $\phi^4$ theory. It was not so crucial to have
terribly high statistics as long as the lattice size was sufficiently large.
Our preliminary results obtained with lower statistics did not differ
considerably from the final ones.
This advantage
could be correlated with our control over the quality of data and the
magnitude of the finite size effects through the bound on the exponent
$\delta$.

\vskip0.5truecm

\noindent{\bf Acknowledgement}
This work is supported by NSF-PHY 92-00148. S.~Kim is supported by
the U.~S. Department of Energy, Contract No. W-31-109-ENG-38.
\vfill\eject

\centerline{\bf References}

\noindent
[1] Y. Nambu and G. Jona-Lasinio, Phys. Rev. {\bf 122}, 345 (1961).

\noindent
[2] D. Lurie and A. Macfarlane, Phys. Rev. {\bf 136}, Bb16 (1964);
K. Tamvakis and G. Guralnik, Nucl. Phys. {\bf B146}, 224 (1978);
A. Hasenfratz, et al., Nucl. Phys. {\bf B365}, 79 (1991);
J. Zinn-Justin, Nucl. Phys. {\bf B367}, 105 (1991).

\noindent
[3] K. Wilson, Phys. Rev. {\bf D7}, 2911 (1973).

\noindent
[4] K. Gawedzki and A. Kupianen, Phys. Rev. Lett. {\bf 55}, 363 (1985);
B. Rosenstein, B. Warr and S. Park, Phys. Rev. Lett. {\bf 62}, 1433 (1989);
C. de Calan, P. Faria da Veiga, J. Magnen and R. S\' en\' eor,
Phys. Rev. Lett. {\bf 66}, 3233 (1991).

\noindent
[5] T. Eguchi, Phys. Rev. {\bf D17}, 611 (1978);
K.-I. Shizuya, Phys. Rev. {\bf D21}, 2327 (1980).

\noindent
[6] M. Aizenman, Comm. Math. Phys. {\bf 86}, 1 (1982);
C. Arag\~ ao de Carvalho, S. Caracciolo and J. Fr\" ohlich,
{Nuc. Phys.} {\bf B215}[FS7], 209 (1983);
M. L\" uscher and P. Weisz, {Nuc. Phys.} {\bf B290}[FS20], 25 (1987);
R. Kenna and C. Lang, Nuc. Phys. {\bf B393}, 461 (1993).

\noindent
[7] A. Casher, Phys. Lett. {\bf 83B}, 395 (1979).

\noindent
[8] S. K. Ma, in {\it Phase Transitions
and Critical Phenomena} Vol.6, eds. C. Domb and M. Green (Academic Press,
London, 1976).

\noindent
[9] K. Wilson and J. Kogut, Phys. Rep. {\bf 12C}, 75 (1974).

\noindent
[10] D. Gross and A. Neveu, Phys. Rev. {\bf D10}, 3235 (1974).

\noindent
[11] B. Rosenstein, B. Warr and S. Park, Phys. Rep. {\bf 205}, 497 (1991).

\noindent
[12]  S.~Hands, A.~Koci\'{c} and J.~B.~Kogut, Phys. Lett. {\bf B273}
(1991) 111.

\noindent
[13] A. Dhar, R. Shankar and S. Wadia, Phys. Rev. {\bf D31}, 3256 (1985).

\noindent
[14] T. Hatsuda and T. Kunihiro, Tsukuba preprint, UTHEP-270.

\noindent
[15] For a review, see T. Appelquist, Yale preprint, YCTP-P23-91.

\noindent
[16] Y. Nambu, in {\it New Trends in Physics}, proceedings of the
XI International Symposium on Elementary Particle Physics,
Kazimierz, Poland, 1988, edited by Z. Ajduk S. Pokorski and A. Trautman (World
Scientific, Singapore, 1989);   V. Miransky, M. Tanabashi and
K. Yamawaki, Mod. Phys. Lett. {\bf A4}, 1043 (1989);
W. Bardeen, C. Hill and M. Lindner, Phys. Rev. {\bf D41}, 1647 (1990).

\noindent
[17] K. Bitar and P. Vranas, FSU preprints, FSU-SCRI-93-127 and
FSU-SCRI-93-130.

\noindent
[18] S.~Hands, A.~Koci\'{c} and J.~B.~Kogut, Ann. Phys. {\bf 224}, 29 (1993).

\noindent
[19] L. K\" arkk\" ainen, R. Lacaze, P. Lacock and B. Petersson,
Saclay preprint, SPhT 93/053.

\noindent
[20] A. Koci\' c and J. Kogut, Illinois preprint, ILL-(TH)-93-21.

\noindent
[21] See, for example,
C.~Itzykson and J.-M.~Drouffe, Statistical Field Theory (Cambridge
University Press, 1989); V. Privman, P.C. Hohenberg and
A. Aharony, in {\it Phase Transitions
and Critical Phenomena} Vol.14, eds. C. Domb and J.L. Lebowitz
(Academic Press, London, 1991).

\noindent
[22]
E. Brezin, J.-C.~Le~Guillou and J.~Zinn-Justin, Phys. Lett. {\bf 47A}
(1974) 285; R. Abe and S. Hikami, Prog. Theor. Phys. {\bf 54} (1975)
1693.

\noindent
[23] B. Freedman and G.~A.~Baker Jr, J. Phys. {\bf A15} (1982) L715.

\noindent
[24] R.~Schrader, Phys. Rev. {\bf B14} (1976) 172;
B.~D.~Josephson, Proc. Phys. Soc. {\bf 92} (1967) 269, 276.

\noindent
[25]
E. Brezin, J.-C.~Le~Guillou and J.~Zinn-Justin, in {\it Phase Transitions
and Critical Phenomena} Vol.6, eds. C. Domb and M. Green (Academic Press,
London, 1976).


\vfill\eject

\magnification=1200

\noindent{\bf Figure captions}

\noindent
1a. The EOS fit Eq.(4.1) to the $16^4$ data.

\noindent
1b. The EOS fit Eq.(4.3) to the $8^4$ simulation data.

\noindent
2a. Same data as in Fig.1a fit to the logarithmic EOS Eq.(2.7a).

\noindent
2b. $8^4$ data plotted following Fig.2a.

\noindent
3a. Susceptibility data on the $16^4$ lattice and power law fits
following Eq.(4.4).

\noindent
3b. Susceptibility data and power law fits on $8^4$ lattice following Fig.3a.

\noindent
4a. $16^4$ simulation data at criticality plotted following Eq.(3.9).

\noindent
4b. $8^4$ simulation data at criticality plotted following Eq.(3.9).
Note the qualitatively {\it different} behavior from 4a.

\noindent
5. Susceptibility data at criticality plotted as in Eq.(4.6).

\noindent
6. $8^4$ simulation data at criticality plotted following scalar field theory.
Finite size effects have distorted the data dramatically.

\noindent
7a. Average action plotted vs. coupling on the $16^4$ lattice.

\noindent
7b. $8^4$ analog of Fig.7a.

\noindent
8a. Specific heat associated with Fig.7a.

\noindent
8b. $8^4$ analog of Fig.8a.

\end

@@@@@@@@@@@@@@@@@@@@@@@@@@@@@@@@@@@@@@@@@@@@@@@@@@@@@@@@@@@@@@@@@@@@@@@@@@@@@@@@

\magnification=800
\pageno=22
\baselineskip=20pt
\parskip=5pt plus 1pt
\parindent=15pt

\centerline{\bf Table I}

\centerline{
Final compilation $16^4$
}
$$\vbox{
\settabs\+\quad$m\backslash\beta$\quad&0.229(3)\quad&0.246(3)
\quad&0.266(3)\quad&0.264(12)\quad&0.279(14)\quad&0.272(45)
\quad&0.287(27)\quad&0.293(25)\quad&\cr
\+\quad$\beta$\quad&\quad $S$ \hfill&\quad $C$ \hfill&\quad $\sigma$
\hfill&\quad $\chi$\hfill&\quad sweeps ($\times 10^4$)\hfill&\cr\bigskip

\+\hfill 0.30 \hfill&\hfill 4.8173(1) \hfill&\hfill 3.8(1) \hfill
&\hfill 1.2123(1) \hfill&  \hfill 0.50(1) \hfill&
\hfill 2 \hfill&  \cr\smallskip

\+\hfill 0.35 \hfill&\hfill 4.5661(3) \hfill&\hfill 3.6(2) \hfill
&\hfill 1.0084(1) \hfill&  \hfill 0.47(2) \hfill&
\hfill 2 \hfill&  \cr\smallskip

\+\hfill 0.40 \hfill&\hfill 4.3255(1) \hfill&\hfill 3.44(8) \hfill
&\hfill 0.82828(5) \hfill&  \hfill 0.55(1) \hfill&
\hfill 3 \hfill&  \cr\smallskip

\+\hfill 0.45 \hfill&\hfill 4.0967(4) \hfill&\hfill 3.438(4) \hfill
&\hfill 0.6613(2) \hfill&  \hfill 0.697(6)\hfill&
\hfill 3\hfill&  \cr\smallskip

\+\hfill 0.50 \hfill&\hfill 3.8820(4) \hfill&\hfill 3.11(15) \hfill
&\hfill 0.4974(2) \hfill&  \hfill 0.988(41)\hfill&
\hfill 3 \hfill&  \cr\smallskip

\+\hfill 0.54 \hfill&\hfill 3.7233(3) \hfill&\hfill 3.01(6) \hfill
&\hfill 0.3592(1) \hfill&  \hfill 1.58(5)\hfill&
\hfill 5 \hfill&  \cr\smallskip

\+\hfill 0.545 \hfill&\hfill 3.7047(2) \hfill&\hfill 3.01(6) \hfill
&\hfill 0.3408(1) \hfill&  \hfill 1.71(4)\hfill&
\hfill 5 \hfill&  \cr\smallskip

\+\hfill 0.55 \hfill&\hfill 3.6864(2) \hfill&\hfill 3.01(6) \hfill
&\hfill 0.3220(1) \hfill&  \hfill 1.90(7) \hfill&
\hfill 5 \hfill&  \cr\smallskip

\+\hfill 0.555 \hfill&\hfill 3.6681(1) \hfill&\hfill 2.84(7) \hfill
&\hfill 0.3025(1) \hfill&  \hfill 2.03(7) \hfill&
\hfill 6\hfill&  \cr\smallskip

\+\hfill 0.56 \hfill&\hfill 3.6504(1) \hfill&\hfill 2.80(7) \hfill
&\hfill 0.2826(1) \hfill&  \hfill 2.24(8) \hfill&
\hfill 6 \hfill&  \cr\smallskip

\+\hfill 0.565 \hfill&\hfill 3.6329(2) \hfill&\hfill 2.79(4) \hfill
&\hfill 0.2619(2) \hfill&  \hfill 2.56(4) \hfill&
\hfill 9 \hfill&  \cr\smallskip

\+\hfill 0.57 \hfill&\hfill 3.6159(1) \hfill&\hfill 2.81(6) \hfill
&\hfill 0.2403(2) \hfill&  \hfill 3.03(8) \hfill&
\hfill 11 \hfill&  \cr\smallskip

\+\hfill 0.575 \hfill&\hfill 3.5995(2) \hfill&\hfill 2.79(7) \hfill
&\hfill 0.2179(2) \hfill&  \hfill 3.61(10) \hfill&
\hfill 10 \hfill&  \cr\smallskip

\+\hfill 0.58 \hfill&\hfill 3.5839(2) \hfill&\hfill 2.73(6) \hfill
&\hfill 0.1945(2) \hfill&  \hfill 4.29(12) \hfill&
\hfill 13 \hfill&  \cr\smallskip

\+\hfill 0.585 \hfill&\hfill 3.5688(1) \hfill&\hfill 2.59(5) \hfill
&\hfill 0.1693(2) \hfill&  \hfill 5.26(11) \hfill&
\hfill 22 \hfill&  \cr\smallskip

\+\hfill 0.59 \hfill&\hfill 3.5544(2) \hfill&\hfill 2.55(4) \hfill
&\hfill 0.1415(3) \hfill&  \hfill 7.18(18) \hfill&
\hfill 20 \hfill&  \cr\smallskip

\+\hfill 0.595 \hfill&\hfill 3.5408(2) \hfill&\hfill 2.66(13) \hfill
&\hfill 0.1091(4) \hfill&  \hfill 12.07(37) \hfill&
\hfill 32 \hfill&  \cr\smallskip


\+\hfill 0.61 \hfill&\hfill 3.5200(1) \hfill&\hfill 1.002(8) \hfill
&\hfill -- \hfill&  \hfill 24.59(1.76) \hfill&
\hfill 13  \hfill&  \cr\smallskip

\+\hfill 0.615 \hfill&\hfill 3.5194(1) \hfill&\hfill 0.971(12) \hfill
&\hfill -- \hfill&  \hfill 14.33(96) \hfill&
\hfill 10 \hfill&  \cr\smallskip

\+\hfill 0.62 \hfill&\hfill 3.51901(1) \hfill&\hfill 0.968(11) \hfill
&\hfill -- \hfill&  \hfill 10.40(65) \hfill&
\hfill 8 \hfill&  \cr\smallskip

\+\hfill 0.625 \hfill&\hfill 3.5186(1) \hfill&\hfill 0.972(5) \hfill
&\hfill -- \hfill&  \hfill 8.02(37) \hfill&
\hfill 9 \hfill&  \cr\smallskip

\+\hfill 0.63 \hfill&\hfill 3.5184(1) \hfill&\hfill 0.978(8) \hfill
&\hfill -- \hfill&  \hfill 6.70(23) \hfill&
\hfill 10 \hfill&  \cr\smallskip

\+\hfill 0.635 \hfill&\hfill 3.5181(1) \hfill&\hfill 0.985(11) \hfill
&\hfill -- \hfill&  \hfill 5.61(18) \hfill&
\hfill 11 \hfill&  \cr\smallskip

\+\hfill 0.64 \hfill&\hfill 3.5178(1) \hfill&\hfill 0.996(10) \hfill
&\hfill -- \hfill&  \hfill 4.93(17) \hfill&
\hfill 11 \hfill&  \cr\smallskip

\+\hfill 0.645 \hfill&\hfill 3.5176(1) \hfill&\hfill 1.022(12) \hfill
&\hfill -- \hfill&  \hfill 4.18(17) \hfill&
\hfill 9 \hfill&  \cr\smallskip

\+\hfill 0.65\hfill&\hfill 3.5173(1) \hfill&\hfill 1.023(12) \hfill
&\hfill -- \hfill&  \hfill 3.82(13) \hfill&
\hfill 9 \hfill&\cr}$$
\vfill\eject

\centerline{\bf Table II}

\centerline{
Final compilation $8^4$
}
$$\vbox{
\settabs\+\quad$m\backslash\beta$\quad&0.229(3)\quad&0.246(3)
\quad&0.266(3)\quad&0.264(12)\quad&0.279(14)\quad&0.272(45)
\quad&0.287(27)\quad&0.293(25)\quad&\cr
\+\quad$\beta$\quad&\quad $S$ \hfill&\quad $C$ \hfill&\quad $\sigma$
\hfill&\quad $\chi$\hfill&\quad sweeps ($\times 10^4$)\hfill&\cr\bigskip

\+\hfill 0.54 \hfill&\hfill 3.7385(4) \hfill&\hfill 2.97(1) \hfill
&\hfill 0.3724(7) \hfill&  \hfill 1.42(5)\hfill&
\hfill 5 \hfill&  \cr\smallskip

\+\hfill 0.55 \hfill&\hfill 3.7036(4) \hfill&\hfill 2.97(1) \hfill
&\hfill 0.3384(7) \hfill&  \hfill 1.65(5) \hfill&
\hfill 5 \hfill&  \cr\smallskip

\+\hfill 0.56 \hfill&\hfill 3.6699(6) \hfill&\hfill 2.88(9) \hfill
&\hfill 0.3030(8) \hfill&  \hfill 1.97(8) \hfill&
\hfill 5 \hfill&  \cr\smallskip

\+\hfill 0.57 \hfill&\hfill 3.6367(2) \hfill&\hfill 2.75(3) \hfill
&\hfill 0.2650(2) \hfill&  \hfill 2.38(3) \hfill&
\hfill 30 \hfill&  \cr\smallskip

\+\hfill 0.575 \hfill&\hfill 3.6206(2) \hfill&\hfill 2.76(4) \hfill
&\hfill 0.2448(2) \hfill&  \hfill 2.78(4) \hfill&
\hfill 30 \hfill&  \cr\smallskip

\+\hfill 0.58 \hfill&\hfill 3.6052(3) \hfill&\hfill 2.69(3) \hfill
&\hfill 0.2242(3) \hfill&  \hfill 3.23(5) \hfill&
\hfill 30 \hfill&  \cr\smallskip

\+\hfill 0.585 \hfill&\hfill 3.5890(2) \hfill&\hfill 2.71(3) \hfill
&\hfill 0.2004(3) \hfill&  \hfill 4.30(12) \hfill&
\hfill 30 \hfill&  \cr\smallskip

\+\hfill 0.59 \hfill&\hfill 3.5739(4) \hfill&\hfill 2.69(5) \hfill
&\hfill 0.1750(8) \hfill&  \hfill 5.93(29) \hfill&
\hfill 30 \hfill&  \cr\smallskip

\+\hfill 0.595 \hfill&\hfill 3.5584(6) \hfill&\hfill 2.57(2) \hfill
&\hfill 0.1447(13) \hfill&  \hfill 8.37(9) \hfill&
\hfill 30 \hfill&  \cr\smallskip

\+\hfill 0.62 \hfill&\hfill 3.5234(1) \hfill&\hfill 1.060(6) \hfill
&\hfill -- \hfill&  \hfill 12.82(19) \hfill&
\hfill 30 \hfill&  \cr\smallskip

\+\hfill 0.625 \hfill&\hfill 3.5220(1) \hfill&\hfill 0.986(8) \hfill
&\hfill -- \hfill&  \hfill 9.92(16) \hfill&
\hfill 30 \hfill&  \cr\smallskip

\+\hfill 0.63 \hfill&\hfill 3.5208(1) \hfill&\hfill 0.942(7) \hfill
&\hfill -- \hfill&  \hfill 7.83(19) \hfill&
\hfill 30 \hfill&  \cr\smallskip

\+\hfill 0.635 \hfill&\hfill 3.5200(1) \hfill&\hfill 0.942(32) \hfill
&\hfill -- \hfill&  \hfill 6.42(11) \hfill&
\hfill 30 \hfill&  \cr\smallskip

\+\hfill 0.64 \hfill&\hfill 3.5193(1) \hfill&\hfill 0.932(34) \hfill
&\hfill -- \hfill&  \hfill 5.40(7) \hfill&
\hfill 30 \hfill&  \cr\smallskip

\+\hfill 0.645 \hfill&\hfill 3.5188(1) \hfill&\hfill 0.886(3) \hfill
&\hfill -- \hfill&  \hfill 4.66(7) \hfill&
\hfill 9 \hfill&  \cr\smallskip

\+\hfill 0.65 \hfill&\hfill 3.5184(1) \hfill&\hfill 0.872(3) \hfill
&\hfill -- \hfill&  \hfill 4.14(10)\hfill&
\hfill 30 \hfill&  \cr\smallskip

\+\hfill 0.655 \hfill&\hfill 3.5180(1) \hfill&\hfill 0.866(4) \hfill
&\hfill -- \hfill&  \hfill 3.71(10) \hfill&
\hfill 30 \hfill&\cr}$$
\vfill
\vfill\eject

\centerline{\bf Table III}

\centerline{
Final Compilation at Criticality ($\beta_c\approx 0.6034$, $16^4$)
}
$$\vbox{\settabs\+\qquad.225\qquad&\qquad1.0230(4)\qquad&\qquad1.0266(4)
\qquad&\qquad1.0290(3)\qquad&\qquad1.0302(3)\qquad&\cr
\+\hfill$m$\hfill&\hfill$\sigma$\hfill&\hfill$\chi$\hfill&
\hfill sweeps $(\times 10^4)$\hfill&\cr
\bigskip

\+\hfill 0.001\hfill&\hfill 0.0962(3) \hfill&\hfill8.87(30)\hfill&\hfill
22\hfill&\cr\smallskip
\+\hfill 0.002\hfill&\hfill0.1212(2)\hfill&\hfill5.87(4)\hfill&\hfill 15
\hfill&\cr\smallskip
\+\hfill0.003\hfill&\hfill0.1389(2)\hfill&\hfill 4.58(13)\hfill&\hfill 16
\hfill&\cr\smallskip
\+\hfill 0.004\hfill&\hfill0.1532(2)\hfill&\hfill3.68(9)\hfill&\hfill16
\hfill&\cr\smallskip
\+\hfill0.005\hfill&\hfill0.1653(2)\hfill&\hfill3.20(8)\hfill&\hfill12
\hfill&\cr\smallskip
\+\hfill0.010\hfill&\hfill0.2093(3)\hfill&\hfill2.11(2)\hfill&\hfill5
\hfill&\cr\smallskip
\+\hfill0.015\hfill&\hfill0.2401(1)\hfill&\hfill1.68(7)\hfill&\hfill5
\hfill&\cr\smallskip
\+\hfill0.020\hfill&\hfill0.2644(1)\hfill&\hfill1.40(2)\hfill&\hfill5
\hfill&\cr}$$
\vfill\eject


\centerline{\bf Table IV}
\centerline{
Final Compilation at Criticality ($\beta_c\approx 0.6055$, $8^4$)
}
$$\vbox{\settabs\+\qquad.225\qquad&\qquad1.0230(4)\qquad&\qquad1.0266(4)
\qquad&\qquad1.0290(3)\qquad&\qquad1.0302(3)\qquad&\cr
\+\hfill$m$\hfill&\hfill$\sigma$\hfill&\hfill$\chi$\hfill&
\hfill sweeps $(\times 10^4)$\hfill&\cr
\bigskip

\+\hfill 0.001\hfill&\hfill 0.1007(9) \hfill&\hfill 16.3(8) \hfill&\hfill
20 \hfill&\cr\smallskip

\+\hfill 0.002 \hfill&\hfill 0.1360(7) \hfill&\hfill 6.9(6) \hfill&\hfill 20
\hfill&\cr\smallskip

\+\hfill 0.003 \hfill&\hfill 0.1559(6) \hfill&\hfill 4.3(5) \hfill&\hfill 20
\hfill&\cr\smallskip

\+\hfill 0.004 \hfill&\hfill 0.1698(5) \hfill&\hfill 3.4(4) \hfill&\hfill 20
\hfill&\cr\smallskip

\+\hfill 0.005 \hfill&\hfill 0.1808(5) \hfill&\hfill 3.1(3) \hfill&\hfill 20
\hfill&\cr\smallskip

\+\hfill 0.006 \hfill&\hfill 0.1905(4) \hfill&\hfill 2.7(2) \hfill&\hfill 20
\hfill&\cr\smallskip

\+\hfill 0.007 \hfill&\hfill 0.1998(4) \hfill&\hfill 2.4(2) \hfill&\hfill 20
\hfill&\cr\smallskip

\+\hfill 0.008 \hfill&\hfill 0.2076(3) \hfill&\hfill 2.2(2) \hfill&\hfill 20
\hfill&\cr\smallskip

\+\hfill 0.009 \hfill&\hfill 0.2147(3) \hfill&\hfill 2.0(2) \hfill&\hfill 20
\hfill&\cr\smallskip

\+\hfill 0.010 \hfill&\hfill 0.2218(3) \hfill&\hfill 1.9(2) \hfill&\hfill 20
\hfill&\cr\smallskip

\+\hfill 0.011 \hfill&\hfill 0.2278(2) \hfill&\hfill 1.9(2) \hfill&\hfill 30
\hfill&\cr\smallskip

\+\hfill 0.012 \hfill&\hfill 0.2341(3) \hfill&\hfill 1.7(1) \hfill&\hfill 30
\hfill&\cr\smallskip

\+\hfill 0.013 \hfill&\hfill 0.2398(2) \hfill&\hfill 1.6(1) \hfill&\hfill 30
\hfill&\cr\smallskip

\+\hfill 0.014 \hfill&\hfill 0.2457(4) \hfill&\hfill 1.5(1) \hfill&\hfill 30
\hfill&\cr\smallskip

\+\hfill 0.015 \hfill&\hfill 0.2506(4) \hfill&\hfill 1.4(1) \hfill&\hfill 30
\hfill&\cr\smallskip

\+\hfill 0.016 \hfill&\hfill 0.2555(3) \hfill&\hfill 1.4(1) \hfill&\hfill 20
\hfill&\cr\smallskip

\+\hfill 0.017 \hfill&\hfill 0.2594(2) \hfill&\hfill 1.3(1) \hfill&\hfill 20
\hfill&\cr\smallskip

\+\hfill 0.018 \hfill&\hfill 0.2639(2) \hfill&\hfill 1.3(1) \hfill&\hfill 20
\hfill&\cr\smallskip

\+\hfill 0.019 \hfill&\hfill 0.2683(2) \hfill&\hfill 1.2(1) \hfill&\hfill 20
\hfill&\cr\smallskip

\+\hfill 0.020 \hfill&\hfill 0.2726(2) \hfill&\hfill 1.2(1) \hfill&\hfill 20
\hfill&\cr}$$
\vfill\eject


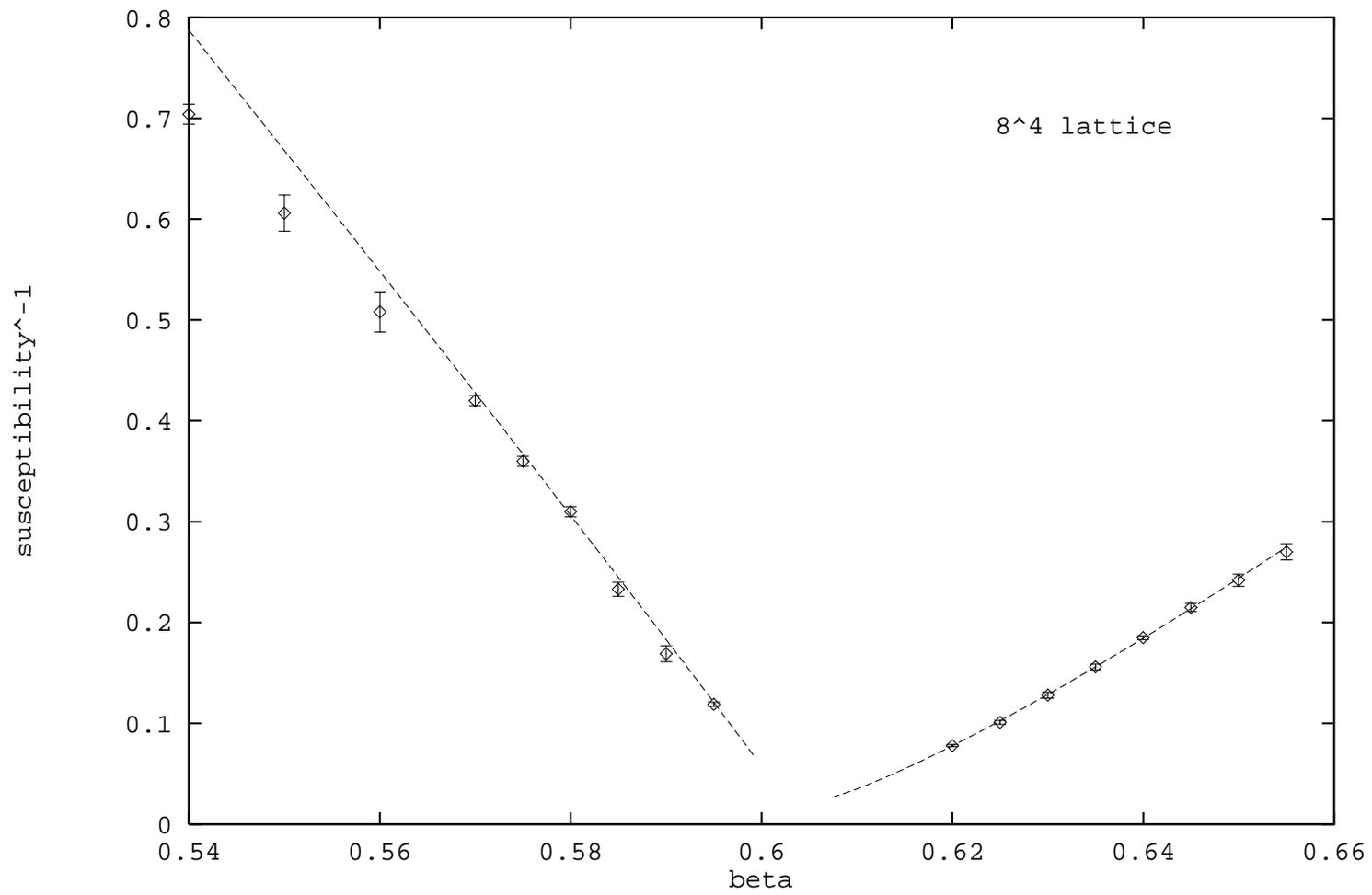
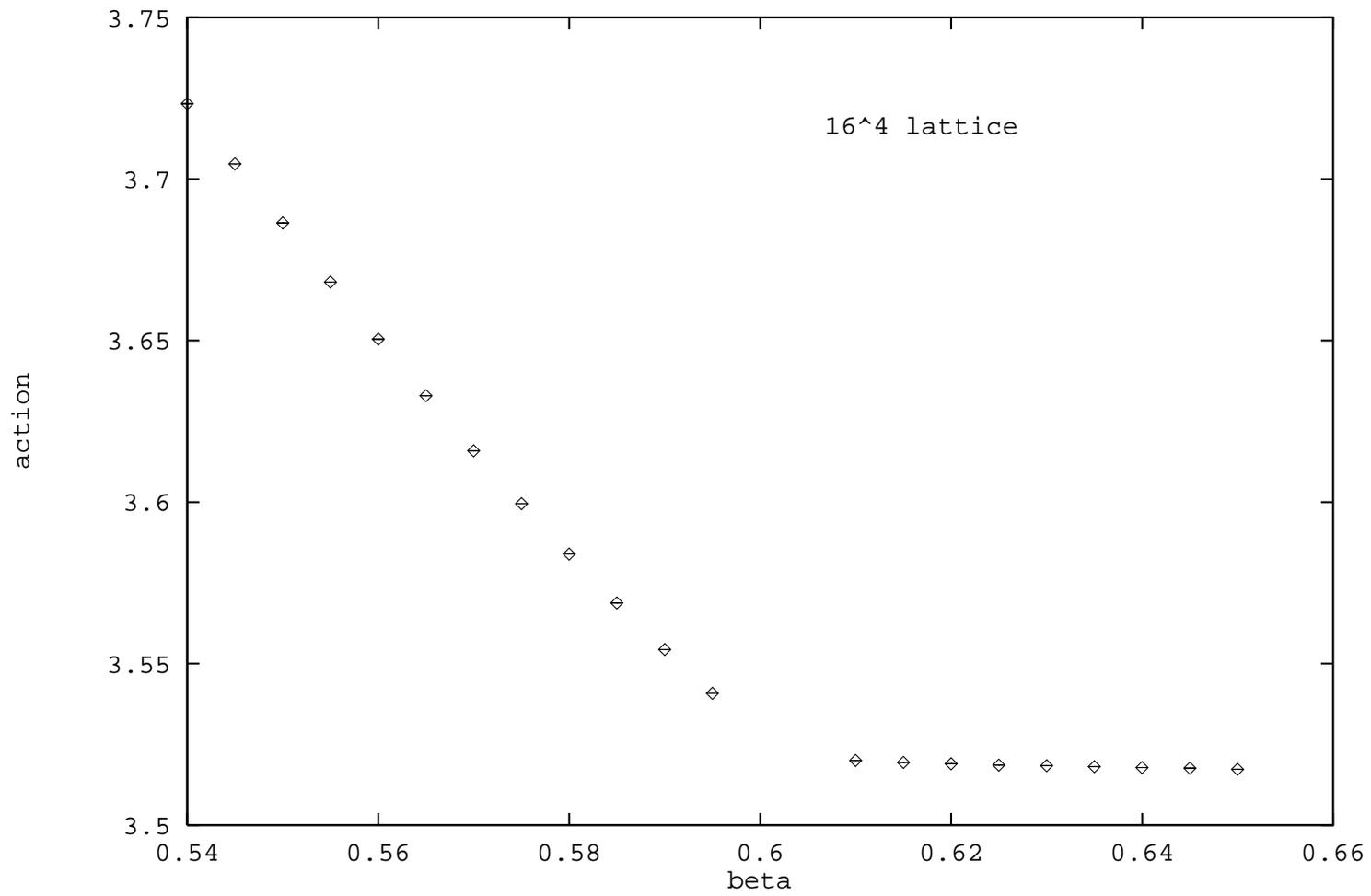
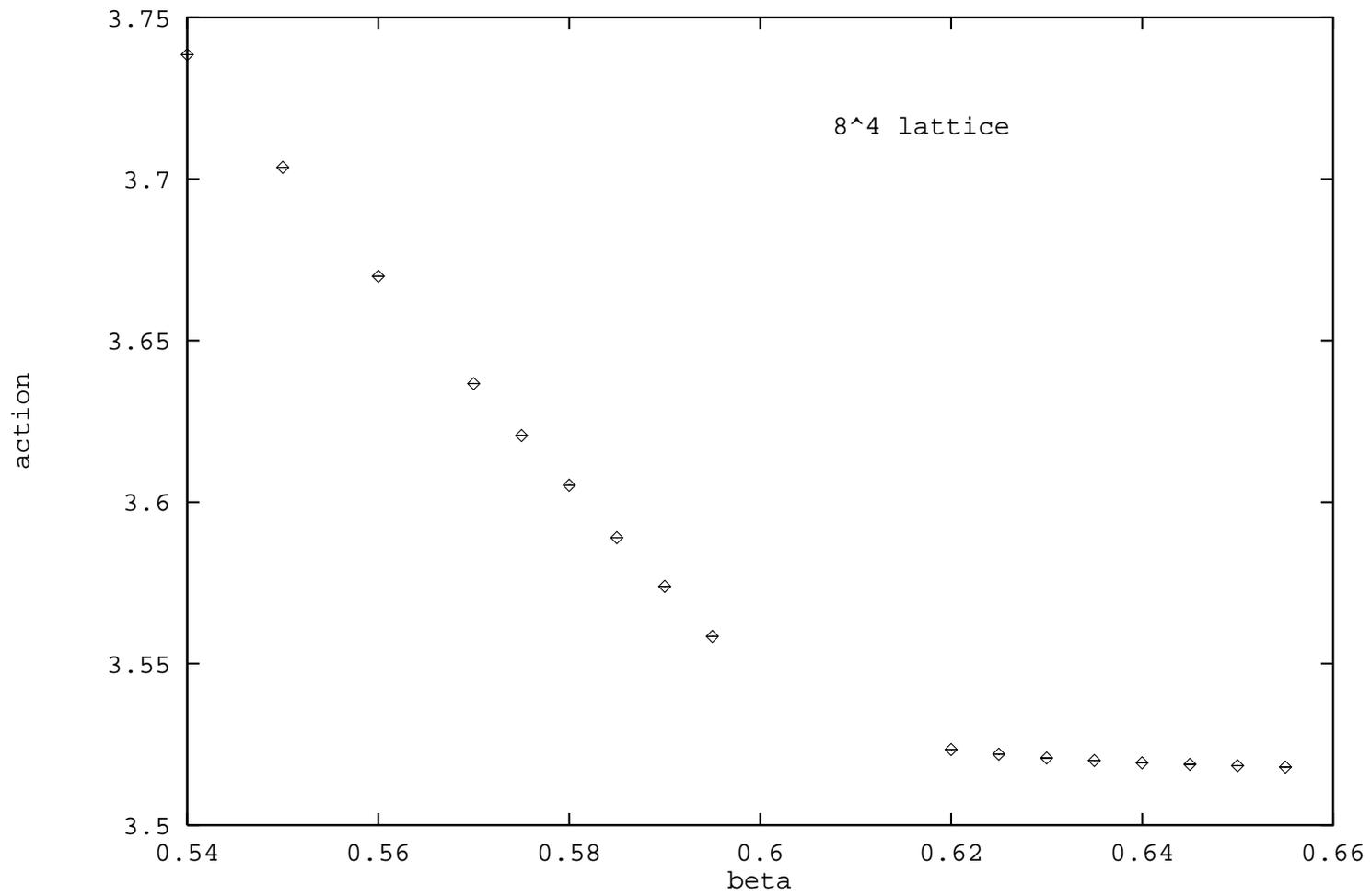
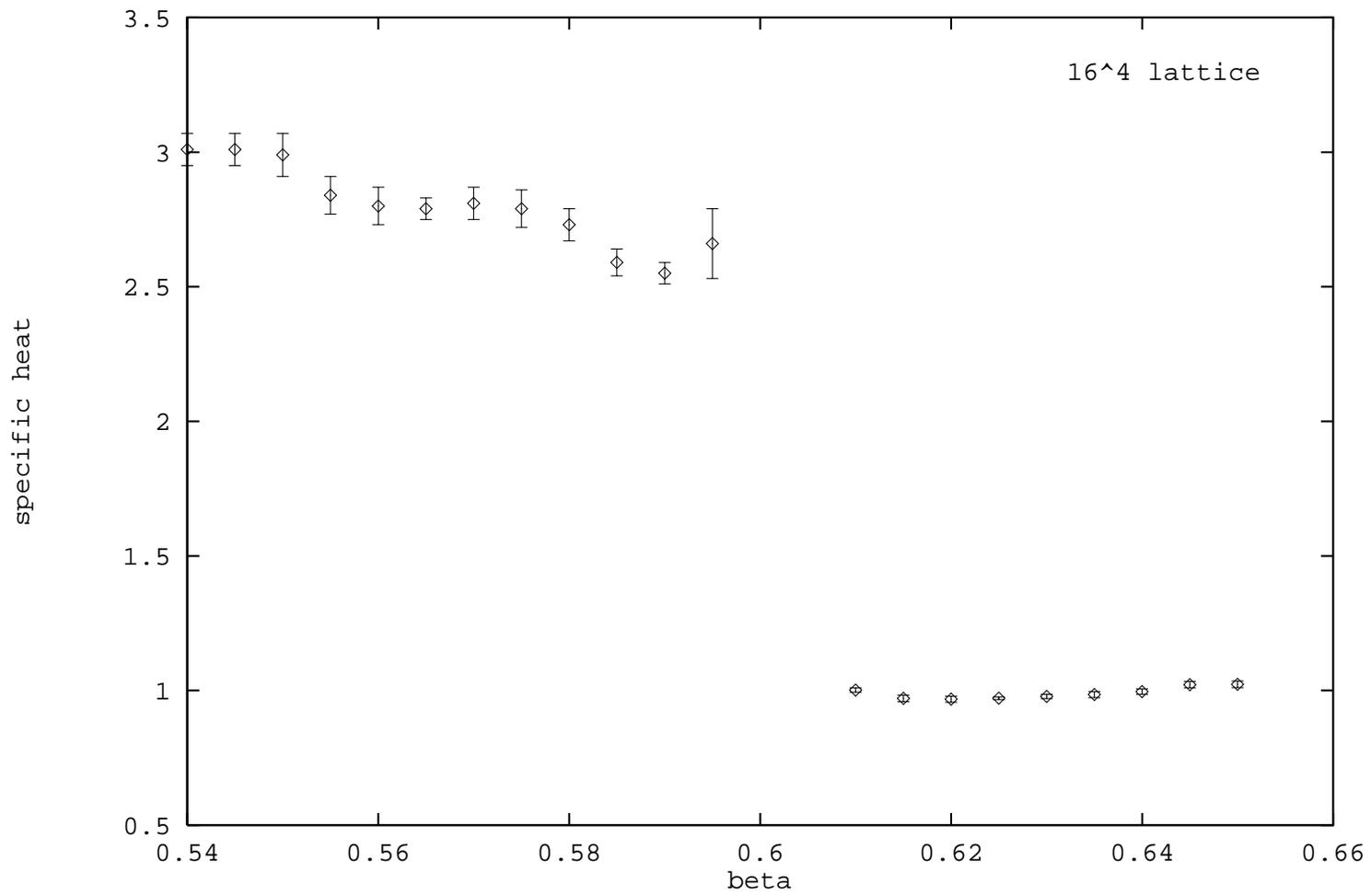
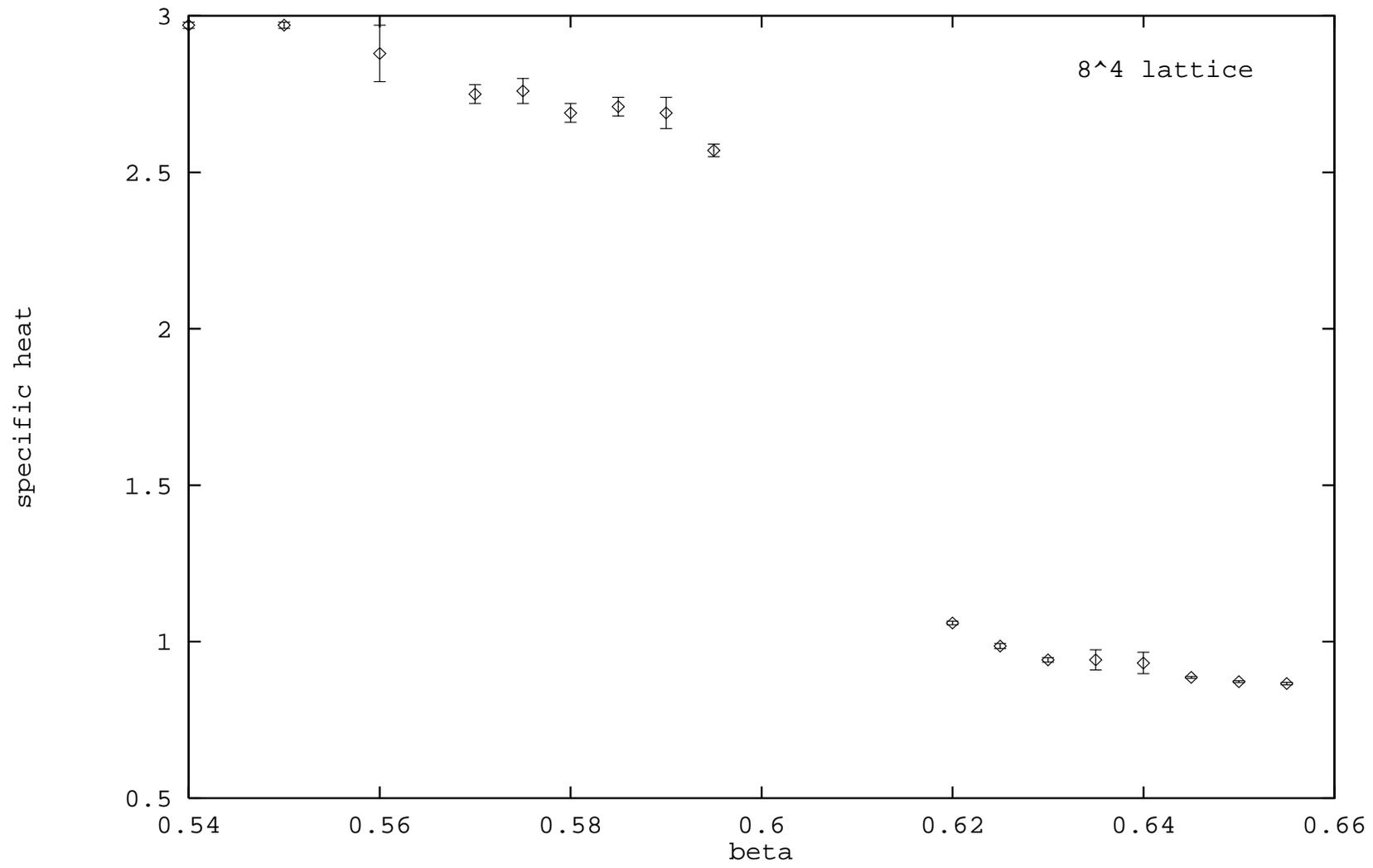
\end